\documentclass[12pt]{article}
\usepackage{graphicx}
\usepackage{scicite}
\usepackage{amsmath}
\usepackage{amssymb}
\usepackage{float}
\usepackage{bm,times}
\usepackage{graphicx}
\newcommand{\be}{\begin{equation}}
\newcommand{\ee}{\end{equation}}
\newcommand{\bea}{\begin{eqnarray}}
\newcommand{\eea}{\end{eqnarray}}

\newenvironment{sciabstract}{%
\begin{quote} \bf}
{\end{quote}}

\newcounter{lastnote}
\newenvironment{scilastnote}{%
\setcounter{lastnote}{\value{enumiv}}%
\addtocounter{lastnote}{+1}%
\begin{list}%
{\arabic{lastnote}.}
{\setlength{\leftmargin}{.22in}}
{\setlength{\labelsep}{.5em}}}
{\end{list}}

\title{Measurement of the Entropy\\ and Critical Temperature of a\\ Strongly Interacting Fermi Gas}

\author{L.~Luo, B.~Clancy, J.~Joseph, J.~Kinast,  and  J.~
E.~Thomas$^{\ast}$\\
\\
\normalsize{$^1$Physics Department, Duke University, Durham, North
Carolina 27708-0305, USA}\\
\\
\ \normalsize{$^\ast$To whom correspondence should be addressed;
E-mail:  jet@phy.duke.edu.}
\\
\normalsize{  }\\
\normalsize{(Submitted 21 November, 2006)}\\
\normalsize{}}
\date{}

\begin{document}

\baselineskip24pt

\maketitle

\begin{sciabstract}

We report a model-independent measurement of the entropy, energy,
and critical temperature of a degenerate, strongly interacting
Fermi gas of atoms. The total energy is determined from the mean
square cloud size in the strongly interacting regime, where the
gas exhibits universal behavior. The entropy is measured by
sweeping a bias magnetic field to adiabatically tune the gas from
the strongly interacting regime to a weakly interacting regime,
where the entropy is known from the cloud size after the sweep.
The dependence of the entropy on the total energy quantitatively
tests predictions of the finite-temperature thermodynamics.

\end{sciabstract}

Strongly interacting Fermi gases are of great interest, as they
exhibit universal thermodynamic behavior, where the properties are
independent of the details of the microscopic
interactions~\cite{OHaraScience,Heiselberg,HoUniversalThermo,ThomasUniversal}.
These gases provide models for testing nonperturbative many-body
theories in a variety of fields from neutron stars and nuclear
matter~\cite{Bertsch,Baker,Heiselberg,Carlson} to quark-gluon
plasmas~\cite{Heinz} and high temperature
superconductors~\cite{Levin}. Hence, thermodynamic experiments on
strongly interacting Fermi gases are of great importance.

In studies of the thermodynamics of these systems, where
thermometry is difficult~\cite{DrummondTemperature}, entropy
measurement plays a central and fundamental role.  We report the
measurement of the entropy $S$ of a strongly interacting Fermi gas
as a function of its total energy $E$. The results yield the
temperature $T$ via the elementary thermodynamic relation
$1/T=\partial S/\partial E$.  Our experiments quantitatively test
recent predictions of the entropy based on microscopic many-body
theory, yield the dependence of the energy on temperature, and
determine the critical temperature for the superfluid transition
without invoking any specific theoretical model.

Strongly-attractive Fermi gases exhibit both fermionic and bosonic
features, and have been studied intensely for several years in
theory~\cite{Randeria,Holland,Timmermans,Levin} and
experiment~\cite{OHaraScience,Jincondpairs,Ketterlecondpairs,Kinast,SalomonBEC,GrimmGap,HuletSpectr}.
Measurements of  the heat capacity~\cite{JointScience} and
collective mode damping versus energy~\cite{KinastDampTemp} reveal
transitions in behavior, which have been interpreted as a
superfluid transition in this system~\cite{JointScience}.
Recently, the observation of vortices~\cite{KetterleVortices} has
provided  a definitive proof of a superfluid phase.  However,
there have been no model-independent studies of the thermodynamic
properties.

A strongly interacting Fermi gas is prepared using  a 50:50
mixture of the two lowest hyperfine states of $^6$Li atoms in an
ultrastable CO$_2$ laser trap with a bias magnetic field of 840 G
just above a broad Feshbach resonance at $B=834$
G~\cite{BartensteinFeshbach}. The gas is cooled  to quantum
degeneracy by lowering the trap depth by a factor of $\simeq
1000$~\cite{OHaraScience}. Following forced evaporation, the trap
depth $U_0$ is recompressed to $U_0/k_B=10\,\mu$K,  which is large
compared to the energy per particle of the gas. Here $k_B$ is the
Boltzmann constant. After this procedure, the initial energy is
close to that of the ground state, as described below.

At the final trap depth, the measured trap  oscillation
frequencies in the transverse directions are $\omega_x=2\pi\times
670$ Hz and $\omega_y=2\pi\times 760$ Hz, while the axial
frequency is $\omega_z=2\pi\times 30$ Hz @ 840 G and
$\omega_z=2\pi\times 32$ Hz @ 1200 G. Note that axial frequencies
differ due to the small change in the trapping potential arising
from the bias magnetic field curvature. The total number of atoms
$N\simeq
 1.3(0.2)\times 10^5$ is obtained from absorption images of the
cloud using a two-level optical transition at 840 G.  The
corresponding Fermi energy $E_F$ and Fermi temperature $T_F$ for
an ideal (noninteracting) harmonically trapped gas at the trap
center are $E_F=k_B T_F\equiv\hbar\,\bar{\omega}(3N)^{1/3}$, where
$\bar{\omega}=(\omega_x\omega_y\omega_z)^{1/3}$. For our trap
conditions, we obtain $T_F\simeq 1.0\,\mu$K.

The total energy per particle, $E$, of the strongly interacting
gas is measured in a model-independent way from the mean square
size in the axial direction~\cite{ThomasUniversal}. In this
strongly interacting regime, the zero energy s-wave scattering
length $a_S$ is large compared to the interparticle spacing, which
is large compared to the range of the two-body interaction, so
that the gas is
universal~\cite{Heiselberg,OHaraScience,HoUniversalThermo}. Then,
the local pressure is $P=2{\cal E}/3$, where ${\cal E}$ is the
local energy density~\cite{HoUniversalThermo,ThomasUniversal}.
Using force balance for a trapping potential $U$, $\nabla
P+n\nabla U=0$, where $n$ is the local density, one then obtains
the total energy per particle $E=3 m\omega_z^2\,\langle
z^2\rangle_{840}\,(1-\kappa)$ or
\begin{equation}
\frac{E}{E_F}=\frac{\langle z^2\rangle_{840}}{z_F^2}  (1-\kappa),
 \label{eq:energy}
\end{equation}
where $\langle z^2 \rangle_{840}$ is  the mean square axial cloud
size measured at 840 G and $m$ is the $^6$Li mass. Here, $z_F^2$
is defined by $3 m\omega_z^2\,z_F^2 \equiv E_F$, and is weakly
dependent on the magnetic field through the trap frequencies. The
correction factor $1-\kappa$ arises from
anharmonicity~\cite{AnharmonicE} in the shallow trapping potential
$U_0\simeq 10\,E_F$ used in the experiments. We find that $\kappa$
varies from 3\% at our lowest energies to 13\% at the highest.

The entropy of the strongly interacting gas at 840 G is determined
using an adiabatic sweep of the magnetic field to a relatively
weakly-interacting regime at 1200 G, where a reference entropy can
be estimated from the mean square axial cloud size $\langle
z^2\rangle_{1200}$. At 1200 G,  $a_S = -2900$
bohr~\cite{BartensteinFeshbach}, and $k_F a_S=-0.75$ for our
shallow trap, with $k_F=\sqrt{2mk_BT_F/\hbar^2}$. At $k_Fa=-0.75$,
we expect that the dependence of the entropy on the cloud size
should be close to that of an ideal noninteracting Fermi gas with
primarily a small mean-field reduction in the ground state cloud
size. This conjecture is supported by the observed ballistic
expansion of the cloud at 1200 G, even at our lowest temperatures,
which shows that the gas is nearly normal. We also find that the
calculated ideal gas entropy differs from a many-body result for
$k_F a_S=-0.75$~\cite{QijinEntropy} by less than 1\% over the
range of energies we studied, except at the point of the lowest
energy, where they differ by 10\%. For this comparison, we
slightly shift the  ground state size of the ideal gas to coincide
with that calculated for $k_F a_S=-0.75$. Hence, the reference
entropy at 1200 G is nearly identical in shape to that for an
ideal gas, and provides a model-independent estimate of the
entropy of the strongly interacting gas.

Ideally, a sweep from 840 G to a magnetic field of 528 G, where
the scattering length vanishes, would produce a noninteracting gas
($k_Fa_S=0$), where the entropy is precisely known. Unfortunately,
adiabatic formation of molecules~\cite{Jinmolec} and subsequent
molecular decay at fields below resonance~\cite{SalomonBEC} cause
unwanted heating for such a downward sweep.

To measure the entropy as a function of energy, we start with an
energy near the ground state and controllably increase the energy
of the gas by releasing the cloud for an adjustable time and then
recapturing it, as described previously~\cite{JointScience}. After
recapture, the gas is allowed to reach equilibrium for 0.7~s. This
thermalization time is omitted for measurement of the ground state
size, where no energy is added.

After equilibrium is established, the magnetic field is either
ramped to 1200 G over a period of 1 s, or the gas is held at 840 G
for 1 s. In either case, after 1 s, the gas is released from the
trap for a short time to increase the transverse dimension of the
cloud for imaging, without significantly changing (less than
0.5\%) the measured axial cloud size.

We find that  the magnetic field sweep is nearly adiabatic, since
the mean square size of the cloud at 840 G after a
round-trip-sweep of 2 s duration is found to be within 3\% of that
obtained after a hold time of 2 s at 840 G. However, we also find
 for our shallow trap that there is a magnetic field and energy
independent heating rate, which causes the mean square size to
slowly increase at a rate of $\dot{\langle z^2\rangle}= 0.024\,
z^2_F/s$,  corresponding to $24$ nK/s in energy units. Since we
desire the energy and entropy just after equilibration, we
subtract $\dot{\langle z^2\rangle}\times 1$ s from the measured
mean square axial dimensions for both the 840 G and 1200 G data.
The maximum correction is 5\% at the lowest energies.

\begin{figure}[tb]
\centerline{\includegraphics[height=4.0in,clip]{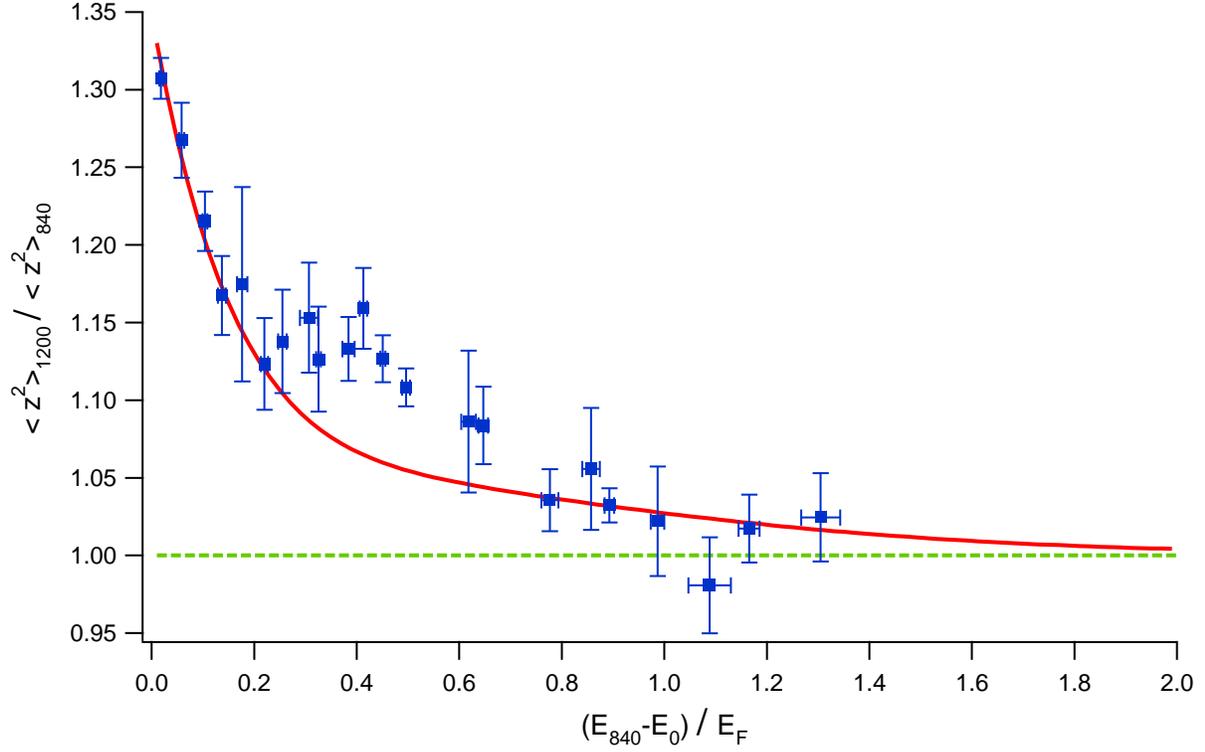}}
\caption{Ratio of the mean square cloud size at 1200 G, $\langle
z^2\rangle_{1200}$ to that at 840 G, $\langle z^2\rangle_{840}$.
 The data is obtained by adiabatically sweeping a bias
magnetic field from 840 G, where the Fermi gas is strongly
interacting, to 1200 G where it is weakly interacting.  $E_{840}$
is the total energy of the strongly interacting gas at 840 G prior
to the sweep,  $E_{0}$ is the ground state energy at 840 G, and
$E_F$ the Fermi energy of a noninteracting gas. The solid line
shows the theoretical prediction based on the calculated
entropies~\cite{QijinEntropy}. The ratio converges to unity at
high energy, as expected (dashed green horizontal line).}
 \label{fig:MeanSq}
 \end{figure}

Fig.~\ref{fig:MeanSq} shows the ratio of the mean square axial
cloud size at 1200 G (measured after the sweep) to that at 840 G
(measured prior to the sweep), as a function of the energy of the
strongly interacting gas at 840 G. The energy at 840 G is directly
measured from the axial cloud size at 840 G using
Eq.~\ref{eq:energy}. The displayed ratio and energy scale are
independent of the atom number and trap parameters. This is
accomplished by measuring the mean square sizes at each field in
units of $z_F^2$ for the given field and atom number. The total
data comprise 900 measurements which have been averaged in energy
bins of width $\Delta E=0.04\,E_F$.

The red solid line shows the predictions obtained by equating the
entropies calculated at 1200 G and near
resonance~\cite{QijinEntropy}. The predicted curve exhibits a
rapid drop followed by a slower decline to unity, in very good
agreement with the data in the low and high energy regions.
However, the data deviate significantly from the prediction in the
region centered near \mbox{$E-E_0\simeq 0.4\,E_F$}, where the
entropy changes behavior as described below.

We note that potential energy has been measured previously in
$^{40}$K~\cite{JinPotential} at a Feshbach resonance and after an
adiabatic sweep to the noninteracting regime. In
Ref.~\cite{JinPotential}, the resulting potential energy ratios
are given as a function of the temperature of the noninteracting
gas. In contrast, by exploiting universality, our cloud size
ratios are referred to the total energy in the strongly
interacting regime, which enables a measurement of $S(E)$ and $T$
for the strongly interacting gas.

For our measurements of $S(E)$, the origin for $S=0$ is determined
by the cloud sizes $\langle z^2\rangle_0$  for the ground states
at 840 G and 1200 G. These sizes are estimated from the data at
the lowest temperatures.  In the harmonic approximation, the
ground state obeys $\langle z^2\rangle_0/z_F^2 = (3/4)\sqrt{\xi}$,
where $\xi\equiv 1+\beta$ is the ratio of the energy per particle
of the strongly interacting gas to that of a noninteracting gas
with the same
density~\cite{Bertsch,Baker,Heiselberg,OHaraScience}. At our
lowest temperatures, including anharmonicity arising from the
gaussian trapping potential $U_{trap}$, we find $\beta_{eff}
=-0.50(0.04)$ at 840 G. For our trap parameters, this corresponds
to $\beta =-0.54 (0.04)$ at 834 G, using the estimate of
Ref.~\cite{ChinSimpleMF}. Our result is in good agreement with
recent measurements based on the axial cloud size, where $\beta
=-0.54(0.02)$~\cite{HuletSpinImbal}, $\beta
=-0.54(+0.05/-0.12)$~\cite{JinPotential} and with recent
calculations, $\beta =-0.56$~\cite{Carlson}, $\beta
=-0.545$~\cite{Strinati},  $\beta=-0.564$~\cite{ChinSimpleMF}.
Using our measured $\beta_{eff} =-0.50$, the ground state energy
per particle for the strongly interacting gas
is~\cite{JointScience} $E_0=(3/4)\sqrt{\xi}\,E_F$, yielding
$E_0=0.53\,E_F$ and $\langle z^2\rangle_0/z_F^2=0.55$ at 840 G.

We can predict the ground state cloud size at 1200 G using the
equation of state at zero temperature. An approximate equation of
state for the chemical potential versus local density, $\mu (n)$,
is given in Ref.~\cite{ChinSimpleMF}. Very good agreement with
quantum Monte Carlo calculations is obtained for negative
scattering lengths, which is the region of interest to us. We
invert the equation of state to find $n(\mu)$, and then using $\mu
=\mu_g -U_{trap}$, we determine the density for a gaussian
potential $U_{trap}$ to include anharmonicity. Normalization to
the number of atoms yields the global chemical potential $\mu_g$
and the mean square cloud size. At 1200 G, where $k_Fa_S=-0.75$,
we find $\langle z^2\rangle_0/z_F^2 =0.69$. Our measurements at
the lowest temperatures yield $\langle
z^2\rangle_0/z_F^2=0.72(0.02)$ at 1200 G, in agreement with the
calculated value. Hence, at both 1200 G and 840 G, we obtain
clouds nearly in the ground state and the corresponding cloud size
ratio $0.72/0.55=1.31$ shown in Fig.~\ref{fig:MeanSq}.

To convert the data of Fig.~\ref{fig:MeanSq} into an entropy
measurement, we calculate the entropy at 1200 G as a function of
the ratio $(\langle z^2\rangle-\langle z^2\rangle_0)/z_F^2$, which
is determined from the axial cloud size data at 1200 G. This
method automatically assures that $S=0$ corresponds to the
measured ground state $\langle z^2\rangle_0$ at 1200 G, and
compensates for small shifts between the calculated and measured
ground state sizes. Then, $S[(\langle z^2\rangle-\langle
z^2\rangle_0)/z_F^2]$ is obtained from   a many-body calculation
at $k_Fa_S=-0.75$, assuming an isotropic gaussian trapping
potential, which automatically corrects for
anharmonicity~\cite{QijinEntropy,QijinThermo}. As discussed above,
nearly identical results are obtained if we assume that the
entropy at 1200 G is that of an ideal Fermi gas in the same
potential.

\begin{figure}[tb]
\centerline{\includegraphics[height=4.0in,clip]{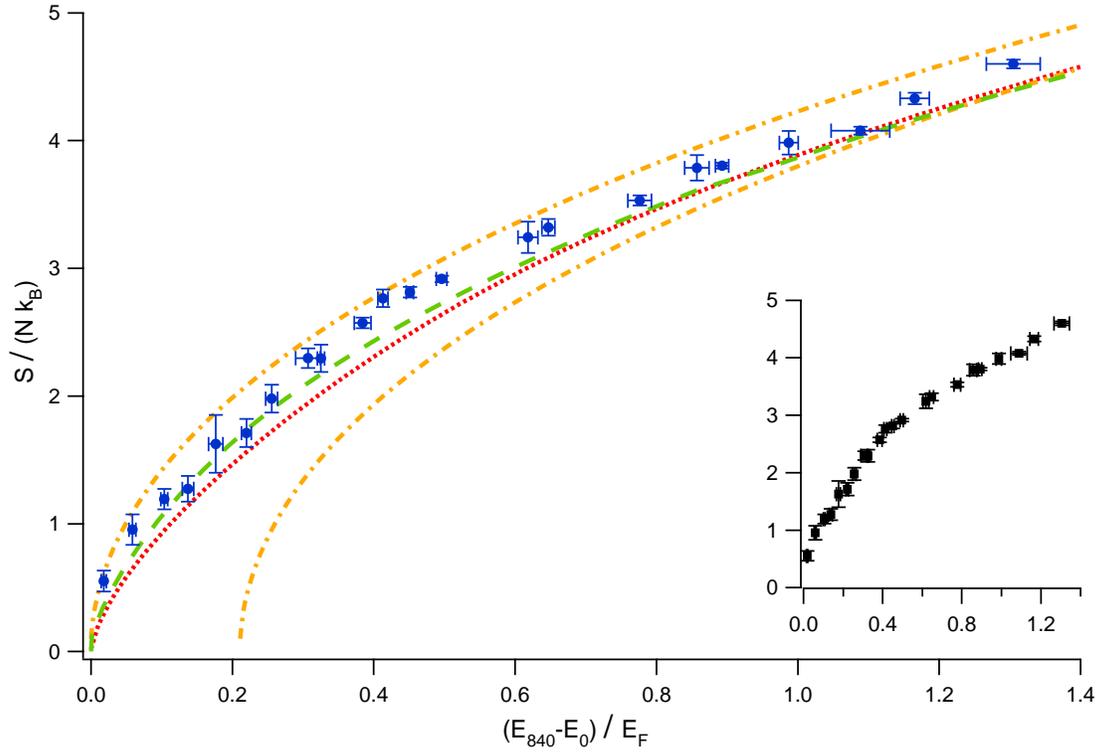}}
\caption{Measured entropy of a strongly interacting Fermi gas at
840 G versus its total energy (blue dots). The entropy is
estimated from the measured cloud size at 1200 G after an
adiabatic sweep of the magnetic field from 840 G. Lower orange
dot-dashed curve-- ideal gas entropy; Upper orange dot-dashed
curve-- ideal gas entropy with the ground state energy shifted to
$E_{0}$; Red dots-- pseudogap theory~\cite{QijinEntropy}; Green
dashes-- quantum Monte Carlo prediction~\cite{BulgacEntropy}.
Inset-- entropy versus energy data showing knee at
$E_c-E_0=0.41\,E_F$.}
 \label{fig:entropy}
 \end{figure}

Fig.~\ref{fig:entropy} shows the entropy (blue dots) of the
strongly interacting gas at 840 G as a function of its energy in
the range $0\leq (E-E_0)/E_F\leq 1.4$. The maximum energy is
restricted to avoid evaporation in the shallow trap, which can
reduce the energy and the atom number during the time of the
magnetic field sweep. The entropy of the strongly interacting gas
differs significantly from that of an ideal gas (lower orange
dot-dash line), which has a larger ground state energy $E_{I0}=
0.75\,E_F$, so that $E_{I0}-E_0=0.22\,E_F$. To compare the curve
shape for the measured entropy to that of an ideal gas, the ideal
gas entropy is also plotted with its energy origin shifted, so
that $S=0$ at $E-E_0=0$ (upper orange dot-dashed line). In
addition, the data are compared to predictions in the resonant
regime based on pseudogap theory~\cite{QijinEntropy,QijinThermo}
(dotted red line) and quantum Monte Carlo methods (dashed green
line)~\cite{BulgacEntropy,BulgacUnitary}.

The temperature is determined in a  model-independent manner from
$1/T=\partial S/\partial E$. This requires parameterizing the
$S(E)$ data to obtain a smooth curve. The simplest assumption
consistent with $S(E=E_0)=0$ is to approximate the data by a power
law in $E-E_0$. However, one expects that below and above the
superfluid transition at a critical energy $E_c$, the power law
exponents will be different. This suggests the simple form,
\begin{eqnarray}
S_<(E)&=&k_B\,a\left(\frac{E-E_0}{E_F}\right)^b
\hspace{0.25in}{\rm for}\hspace{0.125in}
0\leq E-E_0\leq E_c\nonumber\\
S_>(E)&=&S_<(E_c)\left(\frac{E-E_0}{E_c-E_0}\right)^d
\hspace{0.25in}{\rm for}\hspace{0.125in} E-E_0\geq E_c,
\label{eq:entropyscaling}
\end{eqnarray}
where the fit parameters are $a,b,d$ and $E_c$. A fit with this
parametrization yields a $\chi^2$ per degree of freedom $\simeq
1$, a factor of 2 smaller than that obtained by fitting a single
power law to all of the data. However, Eq.~\ref{eq:entropyscaling}
ignores the smooth transition in slope near $E_c$, as required for
continuity of the temperature, since the detailed critical
behavior near $E_c$ is not resolvable in our data.

\begin{figure}[tb]
\centerline{\includegraphics[height=6.0in,clip]{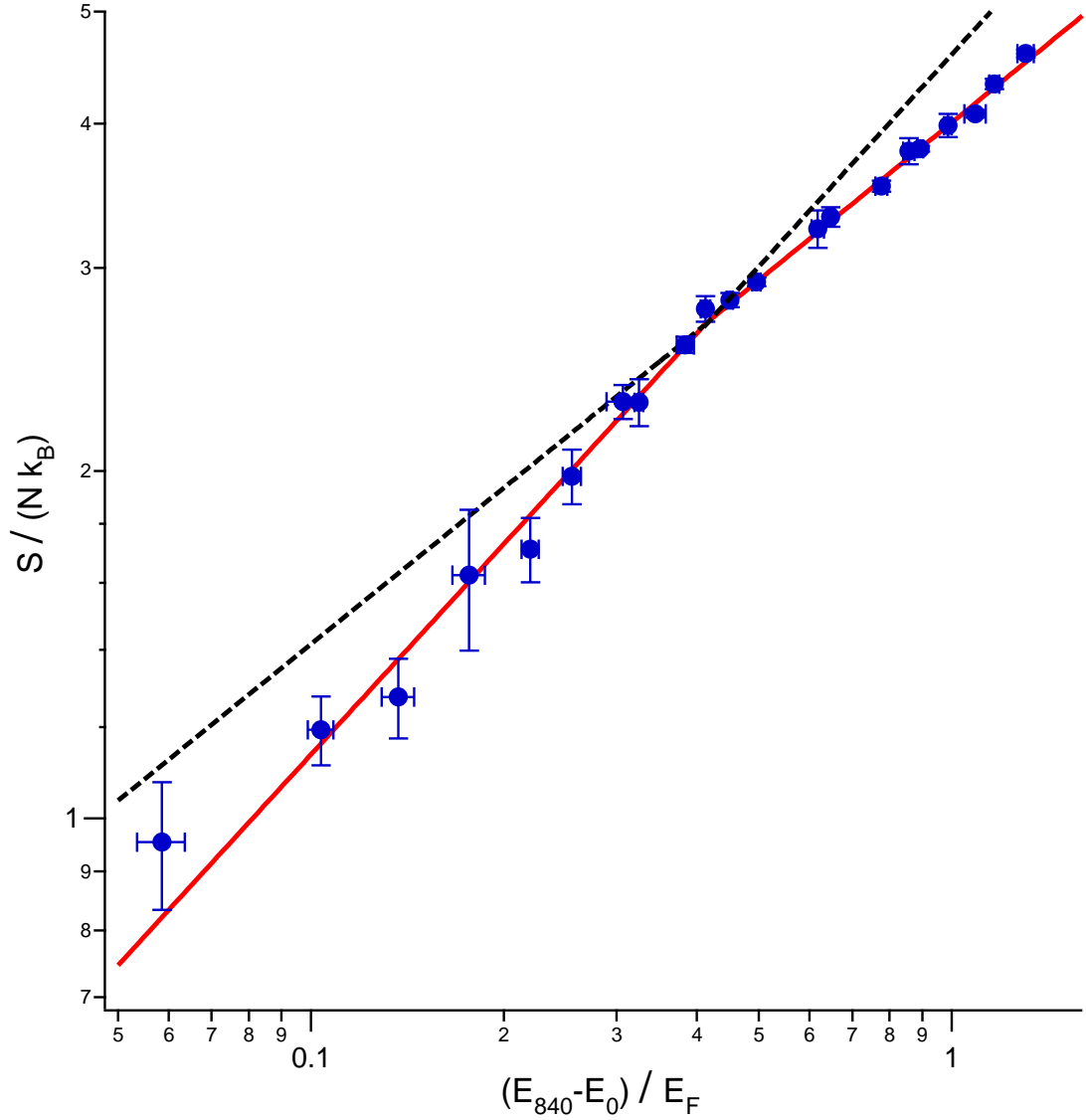}}
\caption{Power law fits for the measured entropy (blue dots) of
the strongly interacting Fermi gas at 840 G versus its total
energy, showing a transition in behavior. Red solid lines show the
fitted power laws below and above $E_c-E_0=0.41\,E_F$. Dotted
black lines show the extended fits. Note that the fit function
does not model the smooth transition in slope near the critical
energy $E_c$, as required for continuity of the temperature.}
 \label{fig:entropylog}
 \end{figure}

Fitting the data of Fig.~\ref{fig:entropy} with
Eq.~\ref{eq:entropyscaling}, the critical energy is found to be
$(E_c-E_0)/E_F=0.41\pm 0.05$, with a corresponding critical
entropy per particle $S_c=2.7(\pm 0.2)\,k_B$. Below $E_c$, the
entropy varies with energy as $S_<(E)=k_B\,(4.6\pm
0.2)\,[(E-E_0)/E_F]^{0.61\pm 0.04}$. Above $E_c$, we obtain
$S_>(E)=k_B\,(4.0\pm 0.2)\,[(E-E_0)/E_F]^{0.45\pm 0.01}$.  We find
that the variances of $a$ and $b$ have a positive correlation, so
that $S(E)$ is determined more precisely than the independent
variation of $a$ and $b$ would imply. The change in behavior near
$E_c$ is shown clearly in the inset of Fig.~\ref{fig:entropy} and
in the log-log plot of Fig.~\ref{fig:entropylog}.

The power law exponent below $E_c$, $b=0.61$, falls between that
of an ideal harmonically-trapped Fermi gas, where a Sommerfeld
expansion at low energy yields $S\propto (E-E_0)^{1/2}$ and that
of an ideal harmonically-trapped Bose-Einstein condensate, where
$S\propto (E-E_0)^{3/4}$. By contrast, above $E_c$, the exponent
$d=0.45$, is close to the result we obtain by fitting a power law
to the entropy of an ideal gas, i.e., $S_I(E-E_{I0})\propto
(E-E_{I0})^b$. In this case, $b=0.485$ for $E-E_{I0}$ below
$0.41\,E_F$ and $b=0.452$ above. This is consistent with the cloud
size ratios shown in Fig.~\ref{fig:MeanSq}, which converge to
unity at higher energies.

The fit parameters from the data can be compared to those obtained
from fits to the theoretical curves shown in
Fig.~\ref{fig:entropy}. The pseudogap
theory~\cite{QijinEntropy,QijinThermo} predicts
$E_c-E_0=0.36\,E_F$, and $S_c=2.16\,Nk_B$. Using
Eq.~\ref{eq:entropyscaling}  to fit the theoretical curve below
the predicted $E_c$, we find $S_<(E) = k_B\,(4.244\pm
0.003)\,[(E-E_0)/E_F]^{0.661\pm 0.005}$. For the quantum Monte
Carlo treatment~\cite{BulgacEntropy,BulgacUnitary}, which predicts
$E_c-E_0=0.32\,E_F$, we find $S_c=2.17\,Nk_B$, and $S_<(E) =
k_B\,(4.35\pm 0.05) \,[(E-E_0)/E_F]^{0.613\pm 0.007}$. Here the
error estimates do not include the error in the theoretical
curves. The small variances indicate that the power law fit
closely approximates the theory, showing that
Eq.~\ref{eq:entropyscaling} is a reasonable parametrization.

The energy versus temperature $E(T)$ is determined from the
derivative of the fit function $S(E)$. For $E\leq E_c$,
\begin{equation}
\frac{E-E_0}{E_F}=\left(\frac{abT}{T_F}\right)^{\frac{1}{1-b}}.
\label{eq:energyvsT}
\end{equation}
From the best fit to the entropy data, where $a=4.6$, $b=0.61$,
$E_c=0.41\,E_F$, we obtain $(E-E_0)/E_F=14\,(T/T_F)^{2.56}$.

We estimate the critical temperature $T_c$ using the measured
value of $E_c-E_0=(0.41\pm 0.05)\,E_F$. Here, we interpret $E_c$
as the critical energy for the superfluid transition. We note that
using $E_0=0.53\,E_F$ yields $E_c=(0.94\pm 0.05)\,E_F$. This value
is consistent with our previous measurements based on the heat
capacity, where we observe a change in behavior at $E =
0.85\,E_F$~\cite{JointScience}, and in collective mode
damping~\cite{KinastDampTemp}, where a plot of the damping rate
versus energy (rather than empirical temperature) shows a change
in behavior near $E=1.01\,E_F$.

Ideally, to obtain $T_c$, the fit $S(E)$ should have a continuous
slope near $E_c$. Since our fit function has different slopes
above and below $E_c$, we approximate the true slope by the
average, as expected for the tangent to a smooth curve. Inverting
Eq.~\ref{eq:energyvsT} yields $T/T_F=0.36\,[(E-E_0)/E_F]^{0.39}$
and $T_{c<}/T_F=0.25$. Similarly, for $E(T)>E_c$, we find
 $T/T_F=0.56\,[(E-E_0)/E_F]^{0.55}$ and $T_{c>}/T_F=0.34$.
Assuming that $2/T_c\simeq 1/T_{c<}+1/T_{c>}$, we find
$T_c/T_F=0.29(0.02)$. Here, the error estimate includes the cross
correlations in the variances of $a$, $b$, $E_c$, and $d$.

The measured critical temperature $T_c/T_F=0.29(0.02)$ can be
compared to our previous estimate of $T_c/T_F=0.27$ from an
experiment with a model dependent temperature
calibration~\cite{JointScience}. Moreover, the result $0.29$ is in
good agreement with predictions for trapped atoms,
0.29~\cite{JointScience}, 0.30~\cite{Torma}, 0.31~\cite{Strinati},
0.30~\cite{BruunViscous}, 0.26~\cite{DrummondTemperature} and
0.27~\cite{BulgacEntropy,BulgacUnitary}.

Transition temperatures also have been predicted for a uniform
gas, $k_BT_c/\epsilon_F^*=0.152$~\cite{BurovskiFermiHubbard} and
$k_BT_c/\epsilon_F^*=0.160$~\cite{ZwergerBECBCS}. These also can
be compared to our measured $T_c$. Here $\epsilon_F^*$ is the
Fermi energy corresponding to the uniform density. By contrast, we
determine the ratio $T_c/T_F$, where $T_F$ is the Fermi
temperature for a noninteracting gas at the center of a harmonic
trap. If we assume that $\epsilon_F^*$ corresponds to the central
density of the strongly interacting gas in our trap, then
$\xi\,\epsilon_F^*=\sqrt{\xi}\,k_BT_F$~\cite{JointScience}. From
this, we estimate $T_c/T_F=k_BT_c/(\epsilon_F^*\sqrt{\xi})$. For
Ref.~\cite{BurovskiFermiHubbard}, we assume $\xi
=0.44$~\cite{Carlson}, and obtain $T_c/T_F=0.23$.
Ref.~\cite{ZwergerBECBCS} calculates $\xi =0.36$ yielding
$T_c/T_F=0.27$.


\begin{thebibliography}{10}

\bibitem{OHaraScience}
K.~M. O'Hara, S.~L. Hemmer, M.~E. Gehm, S.~R. Granade, J.~E.
Thomas, {\it
  Science\/} {\bf 298}, 2179 (2002).

\bibitem{Heiselberg}
H.~Heiselberg, {\it Phys. Rev. A\/} {\bf 63}, 043606 (2001).

\bibitem{HoUniversalThermo}
T.-L. Ho, {\it Phys. Rev. Lett.\/} {\bf 92}, 090402 (2004).

\bibitem{ThomasUniversal}
J.~E. Thomas, A.~Turlapov, J.~Kinast, {\it Phys. Rev. Lett.\/}
{\bf 95}, 120402
  (2005).

\bibitem{Bertsch}
G. F. Bertsch proposed the problem of determining the ground state
of a
  two-component {Fermi} gas with a long scattering length and defined the
  parameter $\xi$.

\bibitem{Baker}
G.~A. Baker{, Jr.}, {\it Phys. Rev. C\/} {\bf 60}, 054311 (1999).

\bibitem{Carlson}
J.~Carlson, S.-Y. Chang, V.~R. Pandharipande, K.~E. Schmidt, {\it
Phys. Rev.
  Lett.\/} {\bf 91}, 050401 (2003).

\bibitem{Heinz}
P.~F. Kolb, U.~Heinz, {\it Quark Gluon Plasma 3\/} (World
Scientific, 2003), p.
  634.

\bibitem{Levin}
Q.~Chen, J.~Stajic, S.~Tan, K.~Levin, {\it Physics Reports\/} {\bf
412}, 1
  (2005).

\bibitem{DrummondTemperature}
H.~Hu, X.-J. Liu, P.~D. Drummond, {\it Phys. Rev. A\/} {\bf 73},
023617 (2006).

\bibitem{Randeria}
C.~A. R.~S. de~Melo, M.~Randeria, J.~R. Engelbrecht, {\it Phys.
Rev. Lett.\/}
  {\bf 71}, 3202 (1993).

\bibitem{Holland}
M.~Holland, S.~J. J. M.~F. Kokkelmans, M.~L. Chiofalo, R.~Walser,
{\it Phys.
  Rev. Lett.\/} {\bf 87}, 120406 (2001).

\bibitem{Timmermans}
E.~Timmermans, K.~Furuya, P.~W. Milonni, A.~K. Kerman, {\it Phys.
Lett. A\/}
  {\bf 285}, 228 (2001).

\bibitem{Jincondpairs}
C.~A. Regal, M.~Greiner, D.~S. Jin, {\it Phys. Rev. Lett.\/} {\bf
92}, 040403
  (2004).

\bibitem{Ketterlecondpairs}
M.~W. Zwierlein, {\it et~al.\/}, {\it Phys. Rev. Lett.\/} {\bf
92}, 120403
  (2004).

\bibitem{Kinast}
J.~Kinast, S.~L. Hemmer, M.~E. Gehm, A.~Turlapov, J.~E. Thomas,
{\it Phys. Rev.
  Lett.\/} {\bf 92}, 150402 (2004).

\bibitem{SalomonBEC}
T.~Bourdel, {\it et~al.\/}, {\it Phys. Rev. Lett.\/} {\bf 93},
050401 (2004).

\bibitem{GrimmGap}
C.~Chin, {\it et~al.\/}, {\it Science\/} {\bf 305}, 1128 (2004).

\bibitem{HuletSpectr}
G.~B. Partridge, K.~E. Strecker, R.~I. Kamar, M.~W. Jack, R.~G.
Hulet, {\it
  Phys. Rev. Lett.\/} {\bf 95}, 020404 (2005).

\bibitem{JointScience}
J.~Kinast, {\it et~al.\/}, {\it Science\/} {\bf 307}, 1296 (2005).
Published
  online 27 January 2005 (10.1126/science.1109220).

\bibitem{KinastDampTemp}
J.~Kinast, A.~Turlapov, J.~E. Thomas, {\it Phys. Rev. Lett.\/}
{\bf 94}, 170404
  (2005).

\bibitem{KetterleVortices}
M.~Zwierlein, J.~Abo-Shaeer, A.~Schirotzek, C.~Schunck,
W.~Ketterle, {\it
  Nature\/} {\bf 435}, 1047 (2005).

\bibitem{BartensteinFeshbach}
M.~Bartenstein, {\it et~al.\/}, {\it Phys. Rev. Lett.\/} {\bf 94},
103201
  (2005).

\bibitem{AnharmonicE}
To estimate the anharmonic correction to the energy, we use the
approximation
  of an isotropic gaussian trapping potential
  $U=U_0[1-\exp(-m\bar{\omega}^2r^2/(2U_0))]$. In this case, we find $\kappa
  =5m\omega_z^2\langle z^4\rangle/(8U_0\langle z^2\rangle)$. For total energies
  above $E_F$, where the spatial distribution is nearly gaussian, one readily
  obtains $\langle z^4\rangle =3\langle z^2\rangle^2$ and $\kappa =
  (5E_F/8U_0)\langle z^2\rangle/z_F^2$.

\bibitem{QijinEntropy}
Q. Chen, pseudogap theory of the entropy, energy and cloud size of
a trapped
  Fermi gas, private communication.

\bibitem{Jinmolec}
C.~A. Regal, C.~Ticknor, J.~L. Bohn, D.~S. Jin, {\it Nature\/}
{\bf 424}, 47
  (2003).

\bibitem{JinPotential}
J.~T. Stewart, J.~P. Gaebler, C.~A. Regal, D.~S. Jin, The
potential energy of a
  $^{40}${K} {F}ermi gas in the {BCS-BEC} crossover (2006).
  ArXiv:cond-mat/06077776.

\bibitem{ChinSimpleMF}
C.~Chin, {\it Phys. Rev. A\/} {\bf 72}, 041601(R) (2005).

\bibitem{HuletSpinImbal}
G.~B. Partridge, W.~Li, R.~I. Kamar, Y.~Liao, R.~G. Hulet, {\it
Science\/} {\bf
  311}, 503 (2006).

\bibitem{Strinati}
A.~Perali, P.~Pieri, G.~C. Strinati, {\it Phys. Rev. Lett.\/} {\bf
93}, 100404
  (2004).

\bibitem{QijinThermo}
Q.~Chen, J.~Stajic, K.~Levin, {\it Phys. Rev. Lett.\/} {\bf 95},
260405 (2005).

\bibitem{BulgacEntropy}
A. Bulgac, quantum Monte Carlo theory of the entropy and energy of
a trapped
  unitary Fermi gas, private communication.

\bibitem{BulgacUnitary}
A.~Bulgac, J.~E. Drut, P.~Magierski, {\it Phys. Rev. Lett.\/} {\bf
96}, 090404
  (2006).

\bibitem{Torma}
J.~Kinnunen, M.~Rodr\mbox{\'{i}}guez, P.~T\mbox{\"{o}rm\"{a}},
{\it Science\/}
  {\bf 305}, 1131 (2004).

\bibitem{BruunViscous}
P.~Massignan, G.~M. Bruun, H.~Smith, {\it Phys. Rev. A\/} {\bf
71}, 033607
  (2005).

\bibitem{BurovskiFermiHubbard}
E.~Burovski, N.~Prokof'ev, B.~Svistunov, M.~Troyer, {\it Phys.
Rev. Lett.\/}
  {\bf 96}, 160402 (2006).

\bibitem{ZwergerBECBCS}
R.~Haussmann, W.~Rantner, S.~Cerrito, W.~Zwerger, Thermodynamics
of the
  {BCS-BEC} crossover (2006). ArXiv:cond-mat/0608282.

\end{thebibliography}

\begin{scilastnote}
\item We thank Ingrid Kaldre for help in constructing the cold
atom source. We are grateful to Qijin Chen and Kathy Levin, U.
Chicago, and Aurel Bulgac and Joaqu\'in E. Drut, U. Washington,
Seattle, for providing calculations of the entropy versus cloud
size in advance of publication. We also thank Jason Ho for many
discussions about entropy and energy measurement, which stimulated
this work. This research is supported by the Chemical Sciences,
Geosciences and Biosciences Division of the Office of Basic Energy
Sciences, Office of Science, U. S. Department of Energy, the
Physics Divisions of the Army Research Office and the National
Science Foundation, and the Physics for Exploration program of the
National Aeronautics and Space Administration.
\end{scilastnote}


\clearpage

\clearpage

\clearpage

\end{document}